# A search for X-rays from five pulsars:

## PSR's 0740-28, 1737-30, 1822-09, 1915+13 and 1916+14

M. A. Alpar[1,2], O. H. Guseinov[1], Ü. Kızıloğlu[1], and H. Ögelman[2]

[1] Physics Department, Middle East Technical University, Ankara 06531, Turkey
[2] Department of Physics, University of Wisconsin-Madison, 1150 University Ave., Madison, WI 53706, USA

**Abstract.** We report observations of PSR's 0740-28, 1737-30, 1822-09, 1915+13 and 1916+14 with ROSAT. In the 0.1-2.1 keV range upper limits to luminosity are derived for power law and blackbody spectra, using a range of $N_H$ estimates. The upper limit to the blackbody luminosity from PSR 1822-09 turns out to be consistent with standard cooling curves. For the other pulsars the upper limits are not restrictive as they are much larger than the luminosities predicted by the models.

## 1. Introduction

Presently there are about 15 rotation-powered pulsars detected in the X-ray region. The X-ray spectra of these pulsars indicate that they are dominated either by a hard power-law like spectrum or a soft blackbody like emission with an effective area commensurate with the neutron star surface. With the exception of the Crab pulsar, most of these detections became possible with the sensitive imaging x-ray telescopes of *EINSTEIN*, *EXOSAT* and *ROSAT*. The standard interpretation of the hard spectra is in terms of the curvature, synchrotron and inverse Compton emission of the relativistic electrons accelerated in the voltage gaps at the boundaries of the open field lines extending to the light cylinder (Cheng, Ho and Ruderman 1986a,b; Harding and Daugherty 1991). For 7 rotation powered pulsars with hard X-ray spectra Ögelman (1994) finds that an empirical relationship of the form $L_x \propto (B\Omega^2)^{2.7}$ can represent very nicely the 7 decades of observed X-ray luminosity with 4 decade variation in B and 2 decade variation in $\Omega$. The pulsars with the soft blackbody like emission, namely Vela, Geminga and PSR's 0656+14, 1055-52 appear to radiate like neutron stars on the initial cooling curve and the bulk of their X-ray luminosity is compatible with the standard cooling scenarios without exotic matter or direct URCA processes (Ögelman 1994).

In addition to the positive detections, there are a number of upper limits to the X-ray flux from radio pulsars. The most complete coverage comes from the *ROSAT all sky survey* where each known radio pulsar was examined for its X-ray emission in the 0.1 to 2.4 keV band (Becker, Trümper and Ögelman 1993). Due to the relatively short exposures ($\sim 300$ s), only 4 new candidates were selected in the above survey; follow-up exposures of these objects are in the analysis stage.

In this paper we describe *ROSAT* observations of five more pulsars, PSR 0740-28, PSR 1737-30, PSR 1822-09, PSR 1915+13 and PSR 1916+14. These pulsars, with characteristic ages $\tau = P/2\dot{P} \sim 2 \times 10^4$ to $\sim 4 \times 10^5$ years, were selected with the intention of looking for initial cooling radiation; PSR 1915+13 was in the field of view of the PSR 1916+14 pointing. In pulsars younger than $\sim 10^4$ yr strong magnetospheric emission swamps the weaker thermal radiation and in pulsars older than $10^6$ yr standard cooling scenarios predict a sharp reduction in the surface temperature when surface photon emission overtakes the neutrino luminosity losses (Nomoto and Tsuruta 1987). Unfortunately, the exposure times were too short to detect a positive signal. In Table 1 we summarize the characteristics of the five pulsars as measured and derived from radio observations (Taylor et al. 1993).

In section 2 we describe the observations. In Section 3 we discuss the implications of the upper limits for the

---

*Send offprint requests to:* Ü. Kızıloğlu
\* *Present Address:* Physics Department, Middle East Technical University, Ankara 06531, Turkey

**Table 1.** Characteristics of the five pulsars as measured and derived from the radio observations (Taylor et al. 1993)

| Pulsar  | Period (sec) | log $\dot{E}$ (erg s$^{-1}$) | log B (gauss) | log age (years) |
|---------|--------------|------------------------------|---------------|-----------------|
| 0740-28 | 0.167        | 35.16                        | 12.23         | 5.20            |
| 1737-30 | 0.607        | 34.92                        | 13.23         | 4.32            |
| 1822-09 | 0.769        | 33.66                        | 12.81         | 5.37            |
| 1915+13 | 0.194        | 34.59                        | 12.08         | 5.63            |
| 1916+14 | 1.181        | 33.71                        | 13.20         | 4.95            |

X-ray luminosities in terms of the adopted $N_H$ values, and distances for these pulsars. In Section 4 the implications of these luminosity limits are discussed both in terms of the thermal cooling models and the empirical magnetospheric luminosity models.

## 2. Observations

All five pulsars were observed with the Position Sensitive Proportional Counter (PSPC) at the focus of the X-ray telescope aboard *ROSAT* in the energy range 0.1-2.4 keV. Detailed descriptions of the satellite, X-ray mirrors, and detectors can be found in Trümper (1983) and Pfeffermann et al. (1986). Table 2 shows the journal of observations including the effective exposure times. The upper limits to the count rates were determined by extracting the photons from a circle of 80″ centered on the expected position of the pulsar and subtracting the background determined from a source free annulus of inner radius 120″ and outer radius 360″ also centered on the source. None of the pulsars showed any significant count rate excess; Table 2 contains the 3σ upper limits.

**Table 2.** Log of the observation for the five pulsars

| Pulsar  | Date             | Exposure (sec) | 3σ Upper Limit (cts s$^{-1}$) |
|---------|------------------|----------------|-------------------------------|
| 0740-28 | 22-23 Aug 1990   | 1772           | 5.6×10$^{-3}$                 |
| 1737-30 | 16-29 Mar 1991   | 3282           | 2.3×10$^{-3}$                 |
| 1822-09 | 30 Mar - 4 Apr 1991 | 5329        | 1.4×10$^{-3}$                 |
| 1915+13 | 9-10 Apr 1991    | 1975           | 26.4×10$^{-3}$                |
| 1916+14 | 9-10 Apr 1991    | 1975           | 8.8×10$^{-3}$                 |

## 3. Conversion of count rate limits to luminosity limits

In order to extract luminosity upper limits from the count rate upper limits we need to know the distance to the pulsar as well as the interstellar column density $N_H$. Furthermore, the luminosity depends on the assumed spectral model. As mentioned in §1, we will use the two spectral models that apply to the already detected pulsars: a sion and a soft blackbody like cooling emission from the neutron star surface.

For the distances, we adopt the values given by Taylor et al. 1993. We compute the $N_H$ values by using the dispersion measure (DM) values and multiplying it with an average conversion factor of 10 hydrogen atoms for each free electron (Seward and Wang 1988). Owing to the large scatter of $N_H$ values for a given DM, we also consider the luminosity limits for $N_H$ values twice as large. Table 3 lists these values for the five pulsars.

**Table 3.** Distance and interstellar column densities adopted in the luminosity upper limit calculations

| Pulsar  | Distance (kpc) | DM (pc cm$^{-3}$) | $N_H$ (cm$^{-2}$) |
|---------|----------------|-------------------|---------------------|
| 0740-28 | 1.9            | 73.8              | (2.3 - 4.6)×10$^{21}$ |
| 1737-30 | 3.3            | 153.0             | (4.7 - 9.5)×10$^{21}$ |
| 1822-09 | 1.0            | 19.9              | (0.6 - 1.2)×10$^{21}$ |
| 1915+13 | 4.1            | 94.8              | (2.9 - 5.9)×10$^{21}$ |
| 1916+14 | 1.6            | 30.0              | (0.9 - 1.9)×10$^{21}$ |

Using the $N_H$ values and distances in Table 3, the energy dependent effective area of the *ROSAT* PSPC and the upper limits to the count rates listed in Table 2, we have calculated the upper limits to the luminosities for the hard power-law like and the soft blackbody like models. For the power-law model we have let the photon spectral index vary between -1 and -3; for the blackbody model we have let the effective temperature vary between 20 and 70 eV. Table 4 lists these luminosity upper limits. For all the pulsars in the list except PSR1822-09, the high $N_H$ values make the blackbody luminosity upper limits very high ($L_{bol} > 10^{35}$ erg s$^{-1}$) and meaningless; they are omitted from Table 4.

**Table 4.** Upper limits to the luminosity (3σ) for the power-law and blackbody models. The low and high values of the luminosities correspond to the low and high values of $N_H$ listed in Table 3. The power-law luminosities are for the energy band 0.1 to 2.4 keV, the blackbody luminosities are bolometric.

| Pulsar  | Power-law $L_x$ Limit (erg s$^{-1}$) | Blackbody $L_{bol}$ Limit (erg s$^{-1}$) |
|---------|----------------------------------------|-------------------------------------------|
| 0740-28 | $(0.9 - 1.4) \times 10^{33}$           | -                                         |
| 1737-30 | $(1.1 - 2.7) \times 10^{33}$           | -                                         |
| 1822-09 | $(1.4 - 2.4) \times 10^{31}$           | $(0.5 - 4.1) \times 10^{32}$              |
| 1915+13 | $(5.3 - 8.0) \times 10^{33}$           | -                                         |
| 1916+14 | $(3.7 - 6.0) \times 10^{32}$           | -                                         |

The measured upper limits to the hard power-law luminosities for all five pulsars (listed in Table 4) are not restricting since they are ~ 2 orders of magnitude above that expected from the empirical magnetospheric X-ray luminosity fit of Ögelman (1994):

$$L_x \simeq 6.6 \times 10^{26} (\frac{B_{12}}{P_s^2})^{2.7} \text{erg s}^{-1}$$

which predicts $4 \times 10^{31}$, $2 \times 10^{31}$, $4 \times 10^{29}$, $8 \times 10^{30}$, and $5 \times 10^{29}$ erg s$^{-1}$ for PSR's 0740-28, 1737-30, 1822-09, 1915+13 and 1916+14 respectively.

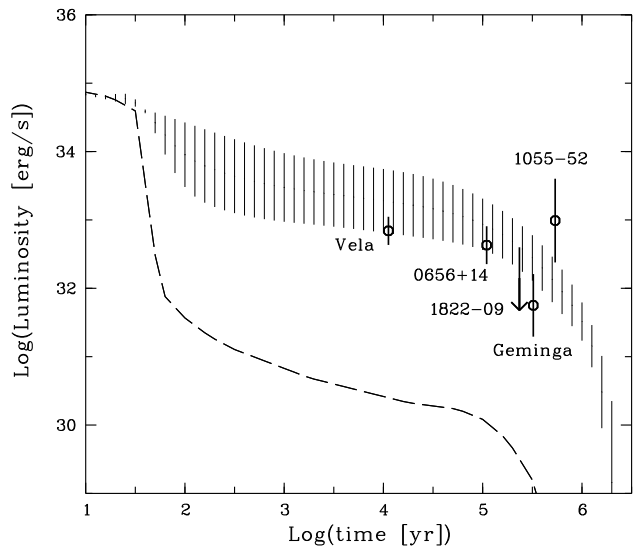

**Fig. 1.** Luminosity versus the dynamic age of the pulsars with soft blackbody like emission (Ögelman 1994) shown together with the upper limit to the similar emission luminosity of PSR 1822-09. The top and bottom of the upper limit arrow corresponds to the limits obtained with the low and high values of $N_H$ listed in Table 3. The theoretical cooling curves of neutron stars are represented by: a) the hatched region corresponding to the range of modified URCA for a 1.3 M$_\odot$ star (top) to direct URCA without various assumptions about the proton and neutron superfluidity (from Page and Applegate 1992); b) the lower dashed curve corresponding to cooling with exotic and direct URCA without neutron superfluidity.

This empirical fit can be taken also as an empirical fit to possible thermal emission from polar cap heating (Cheng and Ruderman 1980 and Arons 1981). This is because the polar cap heating luminosity is fed by magnetospheric processes, and it is therefore reasonable to expect dependence on the voltage parameter, $B/P^2$, at the high temperatures of heated polar caps. The thermal spectrum is similar to a power law spectrum in the 0.1 - 2.4 keV band. To the extent that the same empirical fit can our power law upper limits are too large to yield interesting constraints on the polar cap luminosity. The limits of $L_x/\dot{E}$ is in the range $3 \times 10^{-3} - 10^{-1}$ for these pulsars. Some of the upper limits might give interesting constraints for models of synchrotron nebulae; for comparison $L_x/\dot{E} \sim 5\%$ for the synchrotron nebula around the Crab pulsar.

Neutron stars are expected to dissipate energy as they spin down as the interior will lag the spindown of the crust. According to models of the pinned neutron star superfluid dynamics, the pinned superfluid in the inner crust will lead to energy dissipation at the rate $\dot{E}_{diss} = I_p \bar{\omega}_{cr} |\dot{\Omega}|$ where $I_p \sim 10^{43}$ gr cm$^2$ is the moment of inertia in the pinned superfluid, $\bar{\omega}_{cr}$ the lag in rotation rate between this superfluid and the rest of the star (Alpar et al. 1984, Shibazaki and Lamb 1989, Umeda et al. 1993). This energy dissipation determines the thermal history of older pulsars. For PSR 1737-30, which is a frequently glitching pulsar (McKenna and Lyne, 1990), there could be additional energy dissipation associated with the glitches. If we assume, as we did for polar cap emission, that high temperature thermal emission coming from a limited part of the neutron star surface is represented by our power law fits in the 0.1 - 2.4 keV band, we can derive upper limits to $I_p \bar{\omega}_{cr}$. The lowest upper limit is $I_p \bar{\omega}_{cr} = (1.4 - 2.4) \times 10^{43}$ gr cm$^2$ rad s$^{-1}$ for PSR 0740-28; the others are one or two orders of magnitude larger. In these cases, the upper limits are at least an order of magnitude larger than the upper limit derived from observations of PSR 1929+10 (Alpar et al. 1987).

In terms of the initial cooling luminosities the high $N_H$ values for all except PSR 1822-09 make the blackbody luminosities very high and hence do not yield meaningful limits. On the other hand, for PSR 1822-09 we obtain the $3\sigma$ bolometric luminosity upper limit of $(0.5 - 4.1) \times 10^{32}$ for the $N_H$ values of $(0.6 - 1.2) \times 10^{21}$ respectively. For a pulsar of $2.3 \times 10^5$ yr age this luminosity is a significant upper limit. There are three other pulsars in the same "generation" which have been detected to emit soft blackbody like X-rays with luminosities similar to PSR 1822-09 upper limits, namely PSR 0656+14, Geminga and PSR 1055-52. In Figure 1 we plot the PSR 1822-09 upper limit together with these soft X-ray emitting pulsars and the soft luminosity of the Vela pulsar on a graph of theoretical cooling curves of neutron stars.

As can be seen from the figure, the top of the upper limit arrow of PSR 1822-09 falls on the standard cooling curve. The lower limit (the tip of the arrow), obtained with the lower $N_H$ value, extends even below the range of theoretical predictions without exotic matter or direct URCA cooling. Considering that we have not allowed for errors in the distance, we conclude that the soft X-ray emission limits of this pulsar is compatible with the standard cooling scenarios. It would certainly be worthwhile existing X-ray observatories like *ROSAT* and *ASCA*.

*Acknowledgements.* This research was supported by NASA grant NAGW-2643 and by the Scientific and Technical Research Council of Turkey under the project TBAG-Ü/18-1.